\documentclass[aps,titlepage,12pt]{revtex4-1}
\usepackage{amsfonts}
\usepackage{amsmath}
\usepackage{graphicx}
\usepackage{dcolumn}
\usepackage{bm}
\usepackage{overpic}
\usepackage{booktabs}
\usepackage{color}
\usepackage[justification=raggedright,font={small,sf}, singlelinecheck=false]{caption}
\usepackage{natbib}

\begin{document}
\title{A high-frequency mobility big-data reveals how COVID-19 spread across professions, locations and age groups}

\author{Chen Zhao$^{1,2,3}$, Jialu Zhang$^{1,2,3}$, Xiaoyue Hou$^{1,2,3}$, Chi Ho Yeung$^{4}$}
\thanks{chyeung@eduhk.hk}

\author{An Zeng$^{5}$}
\thanks{anzeng@bnu.edu.cn}

\affiliation{ $^{1}$College of Computer and Cyber Security, Hebei Normal University, 050024 Shijiazhuang, P.R. China\\
$^{2}$Hebei Key Laboratory of Network and Information Security, 050024 Shijiazhuang, P.R. China\\
$^{3}$Hebei Provincial Engineering Research Center for Supply Chain Big Data Analytics and Data Security, 050024 Shijiazhuang, P.R. China\\
$^{4}$Department of Science and Environmental Studies, The Education University of Hong Kong, Hong Kong, P.R. China\\
$^{5}$School of Systems Science, Beijing Normal University, 100875 Beijing, P.R. China }

\begin{abstract}
As infected and vaccinated population increases, some countries decided not to impose non-pharmaceutical intervention measures anymore and to coexist with COVID-19. However, we do not have a comprehensive understanding of its consequence , especially for China where most population has not been infected and most Omicron transmissions are silent. This paper serves as the first study to reveal the complete silent transmission dynamics of COVID-19 overlaying a big data of more than 0.7 million real individual mobility tracks without any intervention measures throughout a week in a Chinese city, with an extent of completeness and realism not attained in existing studies. Together with the empirically inferred transmission rate of COVID-19, we find surprisingly that with only 70 citizens to be infected initially, 0.33 million becomes infected silently at last. We also reveal a characteristic daily periodic pattern of the transmission dynamics, with peaks in mornings and afternoons. In addition, retailing, catering and hotel staff are more likely to get infected than other professions. Unlike all other age groups and professions, elderly and retirees are more likely to get infected at home than outside home.
\end{abstract}


\maketitle

\section*{Introduction}

The SARS-CoV-2, aka COVID-19, pandemic has hit the world since early 2020, and all countries were facing a great challenge in suppressing its transmission and saving lives. Several characteristics of COVID-19 has increased such challenges. First, early variants of COVID-19 have a long incubation period. Second, a substantial proportion of infected individuals may only have minimal symptoms. Before the rapid antigen tests were introduced, infections were mainly verified by PCR tests which are inconvenient and time-consuming. These characteristics lead to silent transmission, i.e. infected individuals are prone to transmit the virus to others without being aware of their own infection. In addition, COVID-19 generally have a high transmissivity, for instance, there are suspected cases of airborne transmission of Omicron~\cite{riediker2022higher, wong2022transmission}. Finally, both hospitalization and fatality by COVID-19 increase with age, which create tremendous pressure on every country's healthcare systems~\cite{ paireau2022ensemble}.

To battle COVID-19, especially with its high silent transmissivity, other than pharmaceutical means such as vaccines and medicines, non-pharmaceutical measures play an important role~\cite{perra2021non, tian2020investigation, willem2021impact}. The ``lockdown" policy which restricts people to stay home is now a common terminology well understood by everyone. Other measures include case isolation and contact tracing, which aim to identify the infected and trace their close contacts and quarantine them~\cite{davis2021contact}; social-distancing to encourage individuals to stay away from each other regardless of being infected or not; travel control to ban travel from local or international origins with infections; closing schools, catering and entertainment premises to avoid gathering~\cite{ lessler2021household, liu2022model, garcia2022public}, etc. Many of these measures aim to prevent silent transmissions without identifying the infected individuals. Finally, some countries adopt herb-immunity and do not impose strict intervention measures when a large portion of their population have been infected or vaccinated.

With this large variety of non-pharmaceutical intervention measures, a comparative analysis which reveals their relative effectiveness can contribute significantly to our battle against COVID-19. Nevertheless, such analysis is difficult, since one cannot fair test these measures in reality while potentially risking lives. Conventional approaches involve employing compartmental models, such as the susceptible-infected-recovered (SIR) model, to analyze these measures; this approach often captures the macroscopic trend, but not the details of the transmission dynamics, such as the location of individual infections, which are crucial for evaluating different interventions~\cite{wu2020nowcasting, ye2020effect, yang2021impact, chang2021mobility, cai2022modeling}. In comparison, agent-based models reveal both the trend and the detailed transmission dynamics~\cite{kerr2021covasim}, for instance, simulation studies with millions of agents were conducted to identify effective intervention measures for France~\cite{thomine2021emerging} and Hong Kong~\cite{zhou2021sustainable} respectively. However, their simulation results can be model artefacts, since they depend crucially on agents' mobility patterns which are only generated by models as these studies lack the empirical mobility data~\cite{vespignani2020modelling}.

As the infected and vaccinated population increases, some countries decided not to impose any non-pharmaceutical measures and to coexist with Covid-19. Despite there is an extensive effort made to understand the effectiveness of intervention measures against COVID-19 and to predict its prevalence, much fewer studies were devoted to reveal the consequence of choosing to live with Covid-19. In particular, some studies show that although vaccination reduces death, hospitalization and symptoms from COVID-19, its effect on reducing transmissions is minimal~\cite{wilder2022vaccine}, which makes silent transmissions more prominent.  This issue is particularly important for China as most population has not been infected and most transmissions are silent due to the Omicron variant.

Thus, to reveal the consequence of coexisting with COVID-19, one relies on a full understanding of the dynamics underlying silent COVID-19 transmissions. Empirical mobility data are essential to reveal the mechanism underlying COVID-19 transmissions but are often difficult to obtain. Hence, some studies used aggregated data such as travel statistics instead of individual mobility patterns to study COVID-19 transmissions~\cite{tian2020investigation, kraemer2020effect, hu2021big}. Other studies used segmented pieces of individual mobility patterns from a small subset of the population to construct mobility tracks for simulations, but results may be sensitive to the model underlying the construction~\cite{aleta2020modelling, aleta2022quantifying, hou2021intracounty}. Other type of data used to study COVID-19 transmissions include self-reported surveys from volunteers ~\cite{allen2020population}, cell phone calls~\cite{vigfusson2021cell} and contacts~\cite{ rudiger2021predicting}, contacts among cruise crews and passengers~\cite{ pung2022using}, etc. Nevertheless, a study with a big data of individual mobility tracks absent in existing studies, can reveal the transmission dynamics of COVID-19 to an extent of completeness and realism not attained so far, and lead to a complete understanding of the dynamics and thus useful insights in our battle against the silent transmission of COVID-19.

In this paper, we use a dataset of 4G communication records between mobile phones and base stations to identify the real mobility tracks of 0.7 million citizens in Shijiazhuang, a city in northern China, in a specific week in 2017, the biggest real mobility dataset employed for the study of COVID-19 to date. Since we aim to reveal the consequence of choosing to coexist with COVID-19 without intervention measures, a dataset before the pandemic is necessary and unique as most other studies used only data during the pandemic which are already influenced by intervention measures. There are over 11500 base stations throughout the city, and position of mobile phones and thus individual users was recorded as the location of the nearest base stations at a high frequency up to every second. We then conduct agent-based simulations using these real mobility data and the empirically inferred infection rate of COVID-19 to reveal the comprehensive transmission dynamics from a few initial infected individuals to a city-wise infection within one week. In addition, one can infer (1) the characteristics of individuals such as age and profession based on their mobility patterns, complying with the city's demographical information, as well as (2) the nature of locations they co-visit with others, complying with the city's geographical information. With these two types of information one can reveal how COVID-19 is transmitted within and across age groups, professions, as well as locations of different nature. This study thus reveals a comprehensive dynamics of COVID-19 transmissions with an extent of realism not achieved in existing studies, and provide useful insights into the choice of coexisting with COVID-19.

\section*{Results}\label{sec2}

\subsection*{Data, model and realistic simulations}

Our big data of empirical mobility tracks is based on 7-day 4G communication records between base stations and mobile phones served by one of the three major service providers between $22^{nd}$ and $28^{th}$ May 2017 in Shijiazhuang, a city in northern China. There are $M=11594$ base stations throughout the city, and the position of an individual is recorded as the location of the nearest base station as long as his/her mobile phones communicate with the base station in 4G. As some phone applications constantly exchange data with back-end servers, the position of individuals can be recorded up to a high frequency of every second.

\begin{figure*}[t!]
\centerline{\includegraphics[width=14 cm]{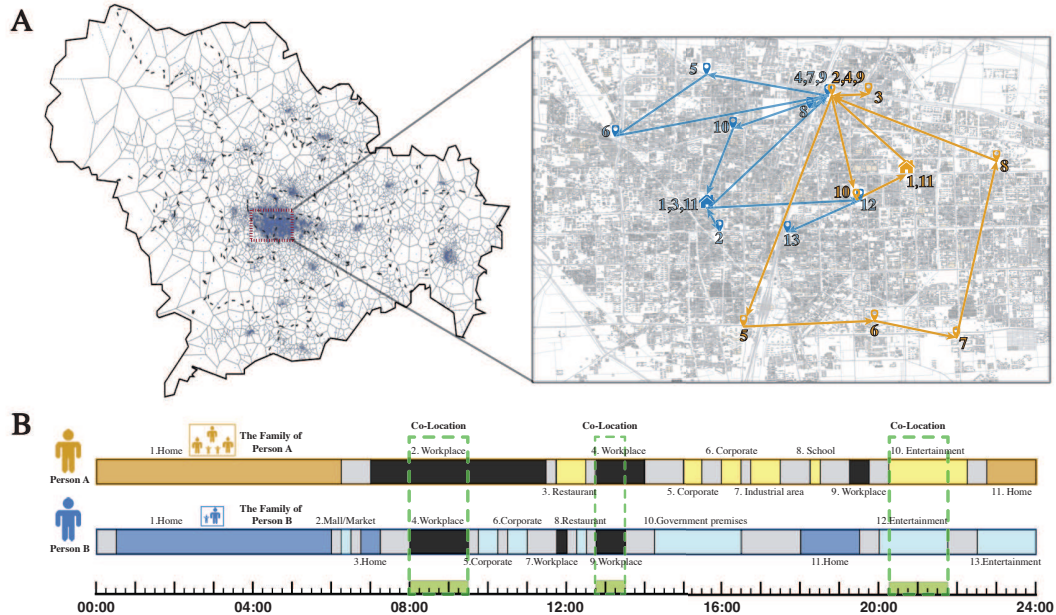}}
\caption{\textbf{Exemplar high-resolution city-wise human mobility tracks and potential silent transmission of COVID-19 via co-location visits.} (A) The map and the distribution of population in Shijiazhuang, and the city center is enlarged and the exemplar mobility tracks from two individuals are shown on the right. (B) Their corresponding sequence of visited locations, numbered according to the location labels in the enlarged map, with the category of each location shown. Blocks in gray correspond to the period when the individuals are ``moving'' Their co-location visits, i.e. they stop at the same location in the same time window, are marked by dashed squares.}\label{fig1}
\end{figure*}

The original data include records from roughly 3 million users out of a total population of 11 millions in the city. To obtain a dataset of valid mobility tracks from active users, we have implemented strict rules to exclude users who do not move at all and those whose data are largely incomplete. Finally, we single out $N=702,477$ valid mobility tracks for simulations (see \emph{Method} for details, and Fig.~S1 in the \emph{Supplementary Note} for the statistics of the dataset). As co-location visits by users
is crucial for transmission of virus, we need to identify meaningful stops in a user's mobility track. We first divide the whole period of the dataset into time windows of 15 minutes as in~\cite{lucchini2021living}, and consider that a user stops in a location if he/she stays there continuously or discontinuously for 10 minutes or more within the time window. Otherwise, if the user did not stay in the same location for more than 10 minutes, we identify the data point as ``moving". A \textbf{\textit{co-location visit}} is defined to occur when two individuals stop in the same location and the same time window.

We follow the formula underlying the COVID-19 Essential Supplies Forecasting Tools (ESFT) introduced by the World Health Organization (WHO)~\cite{world2020covid}, such that the \emph{reproduction number} $R_0$ is given by
\begin{align}
R_0 = D\times\gamma M\times \beta,
\end{align}
where $D$ is the \emph{infectious period}; $M$ is the \emph{average number of co-location visits} with the others per individual per unit time (e.g. day); $\gamma$ is the \emph{fraction of co-location visits which lead to contacts}, i.e. some co-location visits do not lead to contacts, for instance, users with their mobile phones connected to the same base station do not necessary imply a contact, and $\gamma M$ is the number of \textbf{\textit{co-location contacts}}; and $\beta$ is the \emph{probability of infection per co-location contact}. In our simulations, we adopted $R_0=7$ for the Omicron variant of COVID-19~\cite{burki2022omicron}, an infectious period $D = 7$ days~\cite{world2020covid}since we have assumed that the incubation period of Omicron lasts for 1.7  days~\cite{killingley2022safety} and our dataset spans only for 7 days , and finally an average of $M=568$ co-location visits with the others by a single individual per day from our empirical data. In this case, one can estimate the quantity $\gamma\beta\approx 0.002$, which is the probability of infection per co-location contact with the others in a single time window.

By adopting the above model, we study the initial stage of the silent transmission of COVID-19 initiated from $0.01\%$ (i.e. 70 people) of the population (see Supplementary Fig. S2 for the results with fewer initial spreaders). All citizens then move in the city according to their real mobility tracks. We assume that infections start with an incubation period of 1.7 days (see supplementary Fig. S4 for the results with a longer incubation period), and afterwards the infected individuals would be able to infect others~\cite{killingley2022safety}. If a susceptible individual stays at the same location in the same time window (i.e. a co-location visit) with an infected individual passed his/her incubation period, the susceptible may be infected with a probability $\gamma\beta=0.002$ (see Supplementary Fig. S3 for the results of smaller infection probability). Since there may be more than one infected individual at a location, if we denote the number of infected individuals who have passed their incubation period at a location $\alpha$ in the time window at time $t$ to be $n_{\alpha}(t)$, then the probability for a susceptible individual to get infected at $\alpha$ at time $t$ is
\begin{align}
\label{eq_infection}
P_\alpha(t) = 1-(1-\gamma\beta)^{n_\alpha(t)}.
\end{align}
Even if an individual stays in the same location for multiple consecutive time windows, the number of infected individuals may change as time evolves, and the probability for this individual to get infected in a specific time window is given by Eq.~(\ref{eq_infection}) if he/she remains un-infected before the time window. Thus, the real mobility track of individuals determines whether they are in contact with the infected, who can be a stranger in malls, a colleague at the workplace, or a family member at home (see \emph{Method} and later discussions for the inference of individual professions).

We further assume that the infection probability at home is 100 times higher than that outside, for instance, a higher $\gamma$ results from a higher frequency or probability of contacts at home, leading to $\gamma\beta=0.2$ at home (see Supplementary Fig. S5-S6 for the results with other infection probabilities at home). As our data span only for one week, which is shorter than the time needed for recovery from COVID-19, we do not include recovery. We also assume that most infected individuals do not show significant symptoms in a week and the virus spreads silently. This assumption may be more valid for Omicron, since it may lead to milder or even no symptoms compared to other COVID-19 variants~\cite{rajpal2022omicron}. Hence, the mobility behavior of citizens remains un-intervened as they are unaware of the silent transmissions. This makes our dataset suitable since it recorded the mobility behavior in 2017 without the influence of the pandemic, and thus our study is unique since most other studies used only macroscopic datasets during the pandemic.

We first illustrate how our big data of high frequency mobility tracks can lead to simulations of COVID-19 transmission with an extent of completeness and realism not attained in existing studies. As an example, we show in Fig. 1 two individuals who live far away from each other but co-visit some locations at the same time in a day such that silent transmissions may occur between them. Their spatial mobility tracks with various visited locations are shown in Fig. 1A, where both follow a routine schedule like most of us as we can see from their sequence of visited locations in Fig. 1B. In this specific day, these two individuals left home in the morning, and then went to work at the same location, so they had two periods of co-location visits i.e. one in the morning and the other in the afternoon at the work place. After work, they went to the same venue for a concert, which led to their third co-location visit in that day. According to our adopted spreading model, if one of them has been infected with COVID-19 before the day, the other individual would have a probability of $\gamma\beta$ to get infected in each co-location visit in each time window.

\subsection*{Daily periodic transmission dynamics}

\begin{figure*}[t!]
\centering
\includegraphics[width=14 cm]{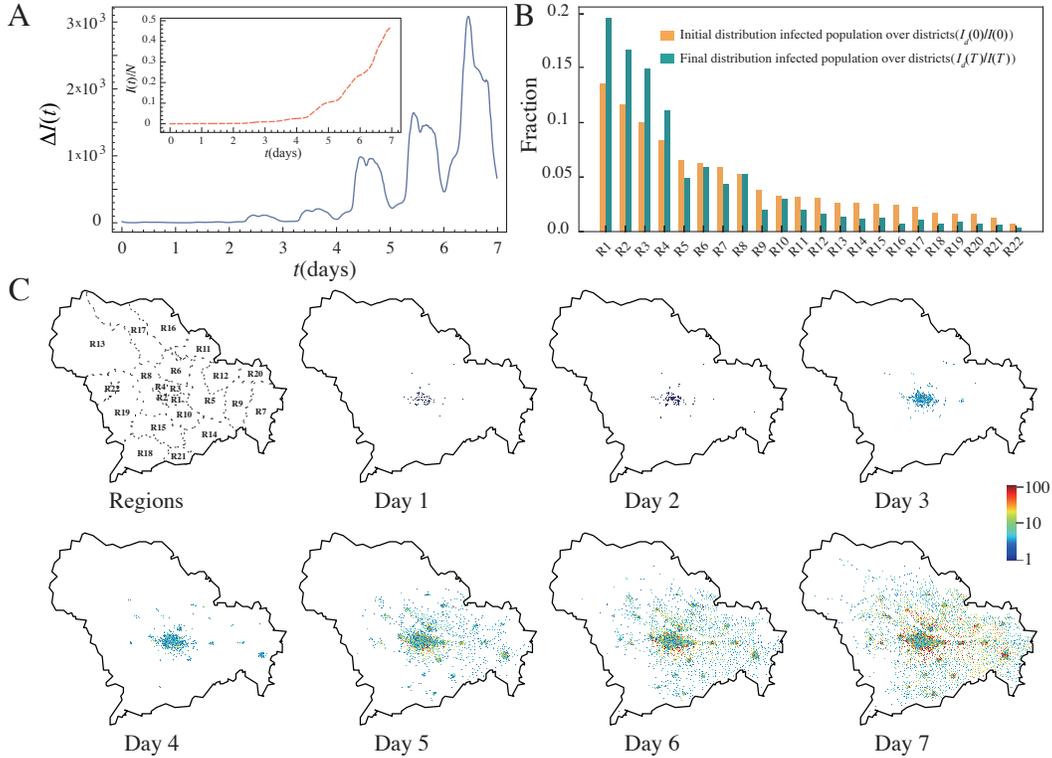}\\
\caption{\textbf{The spatiotemporal patterns of city-wise COVID-19 infection.} (A) The number of new infections $\Delta I(t)$ in the city as a function of time $t$ at a 15-minute interval, given $70$ randomly selected initial spreaders, averaged over $1000 $ realizations. A significant periodic pattern is observed, which is caused by the periodic human mobility behavior. Inset: The corresponding fraction of infected population, i.e. $I(t)/N$. (B) The distribution of the initial and the final infected population over the districts, i.e. $I_d(0)/I(0)$ and $I_d(T)/I(T)$ respectively; districts are numbered as shown in the map in (C). (C) The evolution of the daily spatial pattern of the infected population in the city; the number of infected population in a location is represented by the color of the dot.}\label{fig2}
\end{figure*}

To reveal details of the transmission dynamics difficult to observe without our high-frequency empirical mobility tracks, we show in Fig.~2A the number of new infections $\Delta I(t)$ in the city as a function of time $t$ at a 15-minute interval. According to our analyses, most mobility tracks follow a regular pattern, which are mostly found outside home at different locations in the daytime and stationary at home in the night-time. In general, individuals are more likely to have co-location visits with others and get infected when they visit locations outside home, and less likely to get infected at home, despite the infection rate at home is higher. It is because they came into contact with many others outside home, but only a few family members at home. Therefore, one can observe a significant periodic pattern in $\Delta I(t)$ in Fig. 2A, of which peaks and troughs correspond to daytime and night-time respectively. Interestingly, small troughs are found in between mornings and afternoons, which may correspond to individuals going home for lunch or rest at lunch time. All these results suggest that staying home does help suppressing the silent transmissions of COVID-19, even for a short time like lunch time.

In the inset of Fig.~2A, we show the fraction of infected population as a function of time $t$, i.e. $I(t)/N$, where $I(t) = \sum_{t'=0}^t \Delta I(t')$ is the total number of infections at time $t$. Although there are only $70$ initial spreaders, almost half of the population in our dataset become infected after the week, i.e. a city-wise infection. These transmissions may seem much faster than those in reality; this is because in reality COVID-19 transmissions always come with a high anti-pandemic awareness, preventive and intervention measures, while in our simulations we assume silent transmissions without any interventions. Hence, our results can also serve as a benchmark to demonstrate how COVID-19 spreads given no anti-pandemic awareness nor interventions. However, the fast transmissions observed in our simulations are also dependent on our inferred infection probability $\gamma*\beta$, which may be different from that in reality. Nevertheless, our following major results which compare infections across age groups and professions, locations and districts, etc. are relative to each other and less dependent on the pace of transmissions and thus our estimate of $\gamma*\beta$.

We then go on to examine how transmissions occur in different districts of the city. Such analysis would be difficult even in reality by contact tracing since it is often hard to determine how an individual gets infected~\cite{ davis2021contact}, but would be straightforward in our case by agent-based simulations overlaying a big data of empirical mobility tracks. We first denote $I_d(t)$ as the number of infected individuals at time $t$ who live in district $d$, with $d$ denoting one of the 22 districts in the city (see Supplementary Table 1 for the information of individual districts). We show in Fig.~2B the distribution of the initial and the final infected population over the districts, i.e. $I_d(0)/I(0)$ and $I_d(T)/I(T)$ respectively with $T$ denoting the ending time of the dataset. As the initial infected individuals are randomly selected, their presence should be proportional to the population of a district. As we can see, the final fraction of infected population in a district is not proportional to its initial fraction, i.e. $I_d(T)/I(T) \not \propto I_d(0)/I(0)$, since the infected individuals move across different districts in the city. Interestingly, districts with a large share of initial infected population tend to have an even larger share at last. These districts are mainly regions in the city center with a high population density, and they are also business centers where citizens from other districts come to work or gather in the daytime, causing cross infections. Interestingly, the districts with a small share of initial infected individuals, on the contrary, tend to have an even smaller share of the final infected population. It is because these districts are mostly suburban or rural areas, and their residents are less likely to visit city centers and residents in other areas are also less likely to visit them, which reduces transmissions. In addition, the small population density in these areas may also reduce COVID-19 transmissions since residents have less frequent contacts with others. We show in Fig.~2C the evolution of the spatial pattern of the infected population in the city, which further supports our conjecture that the infections are highly clustered in the city center or regions with a high population density. The results suggest that visiting city centers and meeting strangers from other areas of the city may lead to a high risk of infection.

\subsection*{Profession- and location-dependent transmissions}

\begin{figure*}[t!]
\centering
\includegraphics[width=14 cm]{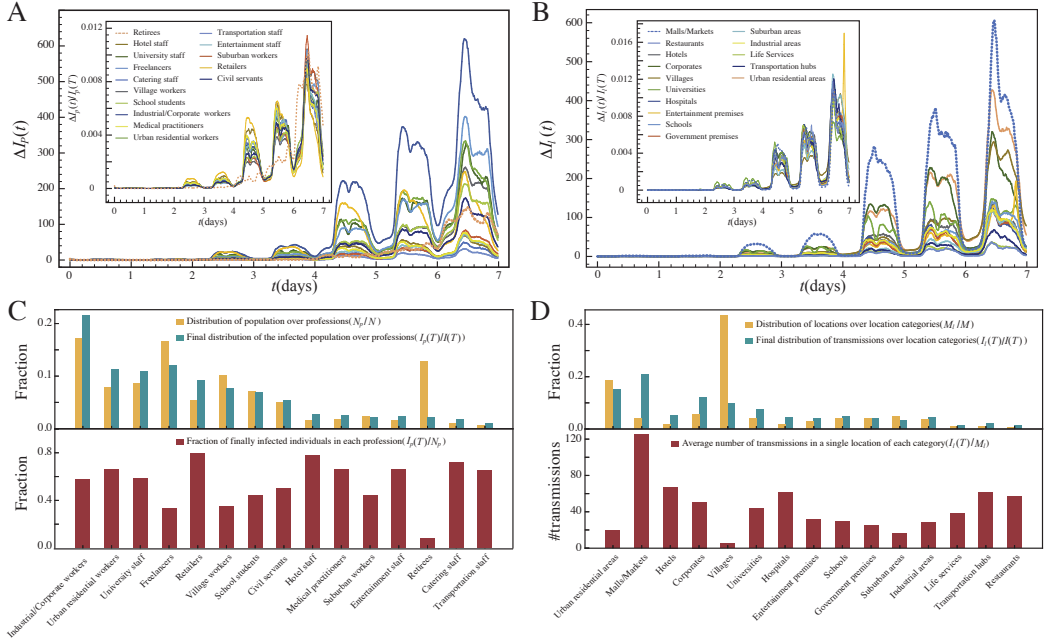}\\
\caption{\textbf{Infected professions and locations.} (A) The number of newly infected individuals of different professions $p$ as a function of time $t$, i.e. $\Delta I_p(t)$. Inset: the fraction of infected population from profession $p$ who get infected at time $t$ , i.e. $\Delta I_p(t)/ I_p(T)$, averaged over $1000 $ realizations. Periodic patterns still exist for different professions, but their infected population is largely different. (B) The number of transmissions in locations of different location category $l$ as a function of time $t$, i.e. $\Delta I_l(t)$. Inset: the fraction of transmissions in location category $l$ which occur at time $t$, i.e. $\Delta I_l(t)/ I_l(T)$. (C) Upper panel: the distribution of population over professions (orange bars), i.e. $N_p/N$, which is proportional to the initial distribution of the infected population over professions $ I_p(0)/I(0)$, and the corresponding final infected distribution (green bars), i.e. $I_p(T)/I(T)$; lower panel: the fraction of final infected individuals normalized by the population size in each profession, i.e. $I_p(T)/N_p$. (D) Upper panel: the distribution of locations over location categories, i.e. $M_l/M$ and the  final distribution of transmissions over location categories, i.e. $I_l(T)/I(T)$; lower panel: the average number of transmissions in a single location of each location category, i.e. $I_l(T)/M_l$.}\label{fig3}
\end{figure*}

Based on the nature of the location of base stations, one can identify the nature of the locations visited by users in their mobility tracks. According to their nature, we classify locations into 15 \emph{location categories} such as malls, schools, etc. With this information, we can infer the profession of individual users (see \emph{Method} for details) and study the relationship between professions and the transmission of COVID-19.

In Fig.~3A, we show how the number of newly infected individuals with different inferred professions increases with time $t$, i.e. $\Delta I_p(t)$ with the subscript denoting inferred profession $p$. An immediate observation is that the periodic pattern still exists for most professions, but the number of infected individuals is largely different across professions as they have a different population size. The top two infected professions are industrial and corporate workers and freelancers, which are also the two professions with the largest population size. To better compare the infection dynamics across professions, we show in the inset of Fig.~3A the fraction $\Delta I_p(t)/I_p(T)$, i.e. the fraction of infected population of profession $p$ who get infected at time $t$; this allows us to compare the periodic pattern of new infections across professions with vastly different population size. One can see that most professions follow the same periodic infection patterns, except retirees who are more likely to be infected in the night-time, possibly caused by their family members who come back home after work.

To compare the likelihood of infection by profession, we show in the \textit{upper panel} of Fig.~3C the distribution of population over professions (orange bars), i.e. $N_p/N$, which is proportional to the initial distribution of the infected population over professions $ I_p(0)/I(0)$ since the initial infected group is randomly selected; we also show the corresponding final infected distribution (green bars), i.e. $I_p(T)/I(T)$. In the \textit{lower panel}, we show the final fraction of infected individuals normalized by the population size in each profession, i.e. $I_p(T)/N_p$. As we can see, despite that retirees take up a large fraction of the population as shown by the orange bars, their final share of infection is small, implying a low infection rate for them as also shown by the red bars in the lower panel. This is an interesting characteristic of the pandemic not revealed in previous studies. Nevertheless, here we assume that all the elderly are living in households with at most 6 family members, and thus transmissions in elderly homes are not considered which can greatly increase the infection rate for the elderly. Unlike retirees who are characterized with the smallest infection rate, professions with the highest infection rate include retailing, catering and hotel management. They are all service professions who are in contact with many strangers every day, and some of these strangers may come from other districts leading to long-distance silent transmissions. If these service providers are infected, they may also act like an infection hub to distribute the virus across different areas of the city, again through the strangers they.

To further understand the reason behind profession-dependent infections, we show in Fig.~3B the number of transmissions in locations of different category as a function of time $t$, i.e. $\Delta I_l(t)$ with $l$ denoting the category of location. This is a quantity which can be measured easily in simulations, but not empirically since transmission are difficult to be identified in reality. As we can see, on top of the periodic patterns, the largest number of transmissions occurs in malls and markets, followed by urban residential areas and corporates. The inset of Fig.~3B shows the fraction of transmissions at locations in category $l$ which occur at time $t$, i.e. $\Delta I_l(t)/I_l(T)$; this again allows us to compare the periodic patterns of new transmissions across location categories with vastly different number of transmissions. One can now see an obvious peak for ``entertainment premises" in the last day of the simulations, which came from a massive gathering in an evening concert (see Fig.~1 for the mobility tracks of two individuals who attended the concert). This shows that large gathering events do pose a high risk of large-scale transmissions.

In Fig.~3D, we show in the \textit{upper panel} the distribution of locations over different categories, i.e. $M_l/M$, where $M_l$ corresponds to the number locations which belong to category $l$; the final distribution of transmissions over location categories, i.e. $I_l(T)/I(T)$, is also shown. In the \textit{lower panel}, we show the average number of transmissions in a single location of each location category, i.e. $I_l(T)/M_l$ . The results suggest that the risk to be infected is highest in malls and markets, hotels and transportation hubs, but lowest in villages and urban residential areas, in consistent with our above findings on profession-dependent transmissions as well as our expectation that locations where people gather are likely for transmission of COVID-19. In this case, stricter intervention measures can be imposed in locations of the high-risk location categories.

Other than profession-related locations, everyone goes home and stay probably a long time there in a day, it is therefore important to examine how often transmissions occur at home. We remark that we do not associate home to any of the location categories, since we define home as the location an individual stayed the longest time over-night instead of based on the nature of the location (see \emph{Method} for details). In Fig.~4A, we compare how the number of transmissions inside and outside home increases with time $t$. As we can see, transmissions occur more frequently outside than at home. The inset of Fig.~4A shows the fraction of transmissions inside and outside home which occur at time $t$, i.e. $\Delta I(t|{\rm home})/I(T|{\rm home})$ and $\Delta I(t|{\rm outside})/I(T|{\rm outside})$ respectively. This again enables us to compare the periodic patterns in spite of the difference in their total number of transmissions. As we can see, both fractions show a periodic pattern but are roughly out of phase, i.e. the peaks of transmissions at home are found roughly in the night-time which are also the minima for transmissions outside.

\subsection*{The dependence of infection on age groups}

\begin{figure*}[t!]
\centering
\includegraphics[width=14 cm]{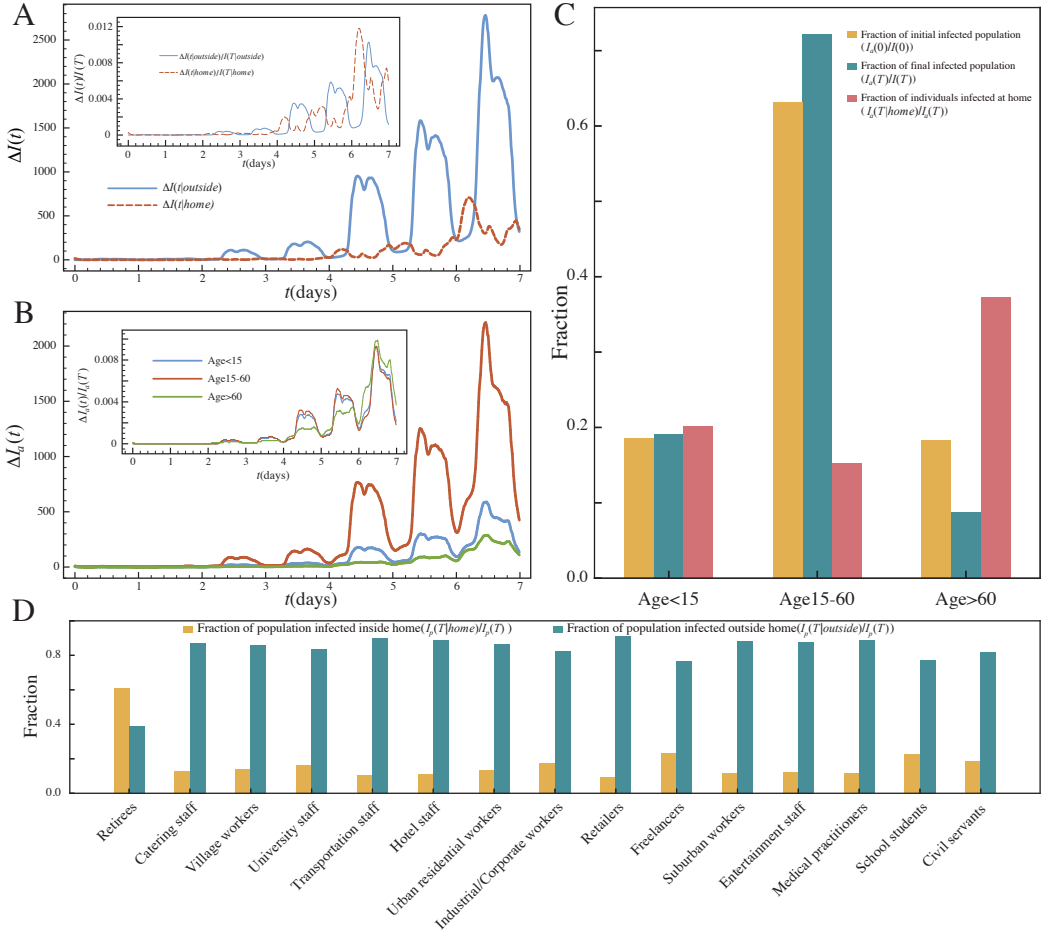}\\
\caption{\textbf{Transmissions inside and outside home and their dependence on age.} (A) The number of new transmissions inside and outside home as a function of time $t$, averaged over $1000 $ realizations. Inset: the fraction of transmissions inside and outside home normalized by the total number of transmissions as a function of time $t$, i.e. $\Delta I(t|{\rm home})/I(T|{\rm home})$ and $\Delta I(t|{\rm outside})/I(T|{\rm outside})$ respectively. (B) The number of new infected individuals as a function of time $t$ in different age groups, i.e. $\Delta I_a(t)$. Inset: the fraction of infections in different age groups which occur at time $t$, i.e. $\Delta I_a(t)/I_a(T)$. (C) The distribution of initial and final infected population across different age groups, i.e. $I_a(0)/I(0)$ and $I_a(T)/I(T)$ as orange and green bars respectively, and the fraction of individuals infected at home in different age groups, i.e. $I_a(T|{\rm home})/I_a(T)$ (red bars). (D) The fraction of population infected inside and outside their home according to their professions, i.e. $I_p(T|{\rm home})/I_p(T)$ and $I_p(T|{\rm outside})/I_p(T)$ respectively.}\label{fig4}
\end{figure*}

Based on the locations individuals visited, their inferred professions and the demographic data from the $7^{th}$ official Census of Shijiazhuang, we can further infer and classify individual users into three age groups, namely (1) under 15, (2) 15 to 60, and (3) above 60 (see \emph{Method} for details). We then go on to investigate how transmissions occur within and across age groups. In particular, both hospitalization and fatality by COVID-19 increase with age, leading to immense pressure on the public healthcare systems, and it is worthwhile to examine how transmissions to the senior population can be suppressed.

In Fig.~4B, we show how the number of newly infected individuals increases with time $t$ in different age groups, i.e. $\Delta I_a(t)$ with $a$ denoting the age group. As the largest fraction of the population falls in the second group with age between 15 and 60, it has the largest share of infected population. We further show in the inset of Fig.~4B the fraction of new infections in different age groups which occur at time $t$, i.e. $\Delta I_a(t)/I_a(T)$. As we can see, senior population with age above 60 evolves differently from those in the other two age groups. Specifically, the senior age group tends to have a higher probability to be infected during the weekend, i.e. the $6^{th}$ and the $7^{th}$ day in our dataset, possibly because their younger infected family members stay at home for a long time or they go out with their family members during the weekend. Fig.~4C further shows the distribution of initial and final infected population across different age groups, i.e. $I_a(0)/I(0)$ and $I_a(T)/I(T)$ respectively, which suggests that in general the senior population is less likely to be infected due to their smaller mobility. The red bars of Fig.~4C also show the fraction of individuals infected at home in different age groups, i.e. $I_a(T|{\rm home})/I_a(T)$. One can see that the senior population are much more likely to be infected at home than younger people.

The above results suggest some specific ways for the senior population to get infected, and are further supported by our profession-dependent analyses. We show in Fig.~4D the fraction of population infected inside and outside their home according to their professions, i.e. $I_p(T|{\rm home})/I_p(T)$ and $I_p(T|{\rm outside})/I_p(T)$ respectively. Since retirees are mainly composed of the senior population, they are the only group who are more likely to be infected at home than outside. As COVID-19 impacts the senior population more severely, to reduce death toll, our results suggest that it is important to avoid the senior population to get infected at home, for instance, to impose prevention measures at home such as wearing masks or to reduce the frequency of high-risk family members visiting or staying with them during the pandemic~\cite{lessler2021household}.

\subsection*{Profession-specific and location-specific source of infection}

\begin{figure*}[t!]
\centering
\includegraphics[width=14 cm]{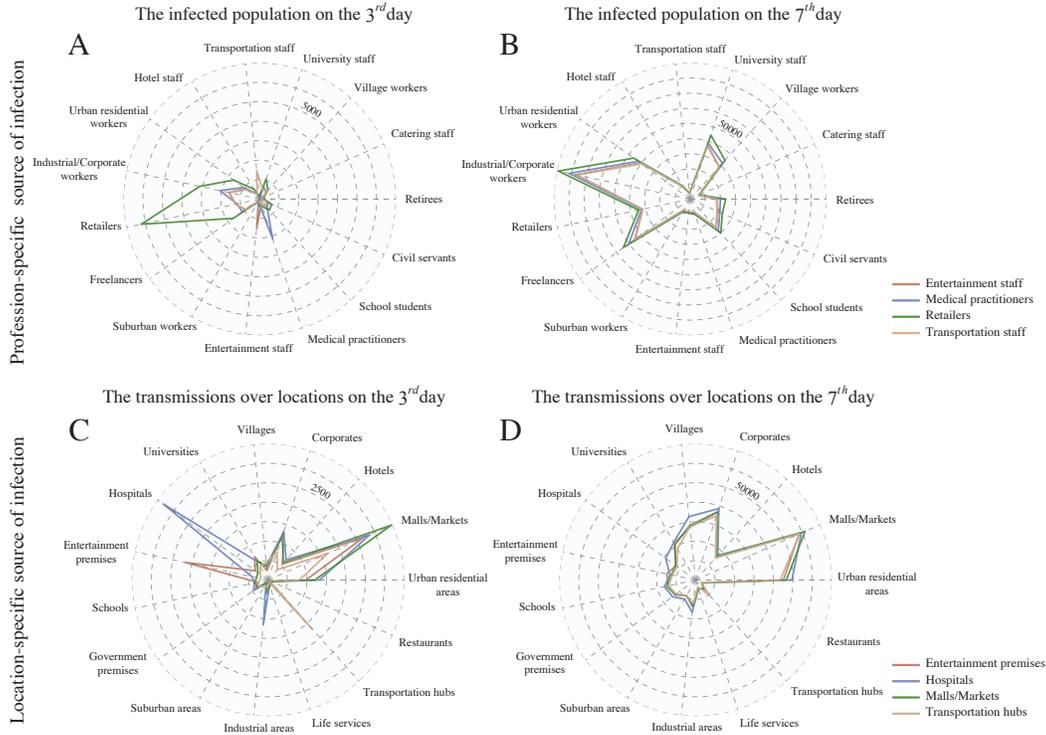}\\
\caption{\textbf{Effect of initial spreaders.} The number of the infected population over professions until (A) the $3^{rd}$ and (B) the $7^{th}$ day of the simulations respectively, i.e. $I_p(\mbox{$3^{rd}$ day})$ and $I_p(\mbox{$7^{th}$ day})$, given that each of the initial infected group falls completely in four different professions. Similar investigation on (C) $I_l(\mbox{$3^{rd}$ day})$ and (D) $I_l(\mbox{$7^{th}$ day})$ for location categories, given that the initial infection starts at locations in four different location categories.}\label{fig5}
\end{figure*}

Finally we study how the final state of infection in the city depends on the initial group of infected~\cite{ odor2021switchover}. Intuitively, the final state depends significantly on the professions and locations of the initial infected group. The radar maps in Fig.~5A and 5B show the total number of the infected population over professions until the $3^{rd}$ and the $7^{th}$ day of the simulations respectively, i.e. $I_p(\mbox{$3^{rd}$ day})$ and $I_p(\mbox{$7^{th}$ day})$, given that each of the initial infected group falls completely in four different professions (see Supplementary Fig. S7 for the results with the initial infected group in other professions). These results suggest that the distribution of infected professions on the $3^{rd}$ day can be substantially different, depending on the initially infected profession. For instance, retailing is the most infected profession given that retailers are initially infected, but less affected if the initial infected group comes from other professions. On the other hand, one can see from Fig. 5B that the final distributions of the infected population initiated by different infected groups look similar, suggesting that the distributions of the infected individuals over professions become more independent of the initial infected groups as time evolves. These results imply that the professions of the initial infected population may have a short-term impact on the distribution of the infected population, but this impact reduces as the pandemic evolves, and ultimately the final state of infection may become independent of the sources.

Similar results can be seen in Fig.~5C and 5D when we investigate the total number of transmissions over location categories until the $3^{rd}$ and the $7^{th}$ day of the simulations respectively, i.e. $I_l(\mbox{$3^{rd}$ day})$ and $I_l(\mbox{$7^{th}$ day})$, given that initial infection starts at locations in four different location categories (see Supplementary Fig. S8 for the results where initial infection starts at other location categories). The results suggest that in the first few days, tracing the sources of infection is important as it affects the professions and locations which are shortly infected. However, in the later stage, source tracing is no longer important as the final infection state becomes independent of the sources.

Finally, we show in the Supplementary Fig. S9 the simulation results where transmissions through both contact and environment are considered. In this case, we first randomly select 70 individuals as the initial infected group and they spread COVID-19 via co-location contacts as in our previous simulations. We then consider four location categories for environmental transmissions, i.e. an individual who visits locations in these four categories would have a probability $0.002$ to be infected on top of the probability via co-location contacts with infected individuals. In other words, visits to these locations may result in infections even if no infected individual is present in these locations in the same time window. As we can see from Supplementary Fig. S9, the results of transmissions are qualitatively similar to those in Fig. 5C and 5D, where one can see a high heterogeneity of infections at the early stage when the environmental spreading occurs at different location categories, but then the states of infection become similar at the later stage.

\section*{Discussion}\label{sec3}

Since more countries decided to coexist with COVID-19, it becomes essential to understand the consequence of such coexistence strategy. From the perspective of China where most population has not been infected and most infections are silent due to the Omicron variant, a comprehensive understanding of the risk of coexisting with COVID-19 is particularly important. In particular, vaccination rate increases in every country but it may make silent transmissions more prominent~\cite{wilder2022vaccine}. Our study served as a first example to simulate the silent transmission of COVID-19 across a community using a big data of more than 0.7 million empirical human mobility tracks without any intervention measures for a week. There were previous studies on COVID-19 transmission using similar agent-based simulations, but they lack the empirical mobility data and have to either generate the mobility patterns of individual agents by models~\cite{thomine2021emerging, zhou2021sustainable}, or construct virtual mobility tracks using segmented pieces of real data~\cite{aleta2022quantifying}. Details including the daily routine of agents, how they co-visit various locations with other agents, how much time they spend in each location, how their mobility patterns depend on their age and professions, etc., are all crucial factors affecting the transmission of COVID-19, but are only modeled or constructed in previous studies, which may lead to a large discrepancy from real mobility patterns and thus a large discrepancy in the results and insights generated. Our study thus represents a call for the use of empirical big data for revealing the realistic transmission dynamics of COVID-19.

With the big data of empirical mobility tracks, we can reveal details for silent transmissions of COVID-19 not observed in previous studies. For instance, our high frequency data record the location of individual users up to every second, such that we can show the number of new infections at a 15-minute interval; this leads us to a daily periodic transmission pattern, which peaks in the mornings and the afternoons as expected since individuals are most active at these times, but an interesting minima in between when some of them go home for lunch or rest. In addition, we also observe less transmissions overnight, suggesting staying home does mitigate COVID-19 transmissions. If individual mobility patterns are generated by models instead of obtained from empirical data, the results would be largely sensitive to model formulation and we are not sure whether such interesting results are simply model artefacts or real phenomenon. This again suggests that agent-based simulation using empirical big data can reveal realistic transmission dynamics and thus useful insights that help in our battle against COVID-19.

Indeed, the largest advantage brought by our big data of mobility tracks is not limited to the high frequency macroscopic trends, but the very details such as how transmissions depend on age, profession, and location. They are obviously crucial factors influencing transmissions, but to reveal their relationship with transmissions one need to model how mobility depends on them which are difficult and are yet to be examined in previous studies. In our case, with the empirical data of mobility tracks, we can reveal the dependence of transmission mechanism on age, profession and location relatively easily by inferring the characteristics of individuals and locations. This again gives us useful insights such as the professions and the nature of locations at which transmissions are more likely to occur, or the common ways senior citizens get infected, which are hidden mechanisms difficult to be revealed even in reality. In particular, we found that retailing, catering and hotel management are the professions prone to COVID-19 infection, while retirees and elderly are less likely to be infected due to their limited mobility. As for the nature of locations, the largest number of transmissions occurs in malls and markets, followed by  urban residential areas and corporates. Moreover, staying home does help suppressing COVID-19, even for a short time and with a larger infection rate at home.

A follow-up study with a great potential to fully utilize our big data of mobility tracks is to reveal the effectiveness of different non-pharmaceutical measures in suppressing COVID-19. Early studies which investigate aggregated data such as travel statistics have already shown correlations between the implementation of non-pharmaceutical measures, such as travel control, with the reported number of COVID-19 infections, but comprehensive individual mobility patterns are not studied~\cite{tian2020investigation, kraemer2020effect}. Large-scale agent-based simulations are also used for this purpose, but as we have mentioned before, agents' mobility tracks in these studies are generated by models~\cite{thomine2021emerging, zhou2021sustainable}. We expect that with the big data of mobility tracks, one can investigate intervention measures such as lockdown, isolation, contact tracing, quarantine, travel control, etc. Nevertheless, there are also shortcomings of studying intervention measures with real mobility tracks as they are real data at a time without the pandemic and models are required if we would like to simulate individual mobility subject to these measures. However, the extent of modeling in this case would be less than those without empirical data.

COVID-19 has impacted every aspect of our daily life since early 2020~\cite{bavel2020using}, and its characteristics make it prone to silent transmissions and thus difficult to be identified in the community before some infected individuals show up with obvious symptoms. There is no easy way to reveal how the virus is transmitted throughout the community, and we believe that our approach of large-scale agent-based simulations overlaying a big data of individual mobility tracks is a promising one with a sufficient extent of completeness and realism not attained in existing studies. We hope that the insights generated in our study would contribute to our battle against COVID-19.


\begin{table*}[t!]
\caption{The average, median, $25^{th}$ and $15^{th}$ percentile of daily total traveling distance and the radius of gyration estimated from the mobility tracks of all users.}
\centering
\begin{tabular}{|c|c|c|c|c|}
\hline
& Average & Median & $25^{th}$ percentile & $15^{th}$ percentile \\
\hline
Daily total traveling distance (meters) & 17485 & 10331 & 2829 & 791 \\
Radius of gyration (meters) & 2435 & 1423 & 497 & 213 \\
\hline
\end{tabular}
\label{tab_distance}
\end{table*}

\begin{table}[t!]
\caption{The statistics of the number of members per household in Shijiazhuang based on the $7^{th}$ Census in 2022.}
\centering
\begin{tabular}{|c|c|}
\hline
Number of household members & Percentage \\
\hline
1 & 0.1243 \\
2 & 0.2716 \\
3 & 0.2316 \\
4 & 0.1902 \\
5 & 0.097 \\
6 & 0.0943 \\
\hline
\end{tabular}
\label{tab_household}
\end{table}

\section*{Materials and methods}

\textbf{Dataset of mobility tracks.}
Our original data is composed of 7-day 4G communication records between base stations and mobile phones served by one of the three major service providers between $22^{nd}$ and $28^{th}$ May 2017, from close to 3 million users out of a total population of 11 millions in Shijiazhuang, a city in northern China. There are over 11500 base stations throughout the city, and the position of an individual is recorded as the location of the nearest base station as long as his/her mobile phones communicate with the base stations in 4G. As some phone applications constantly exchange data with back-end servers, the position of individuals can be recorded up to a high frequency of every second. We then divide each single day into $96$ time windows, each with a duration of 15 minutes, and consider a user stops in a location if he/she stays there continuously or discontinuously for 10 minutes or more within the 15-minute time window. Otherwise, if the user does not stay in the same location for more than 10 minutes within a time window, we identify the data point as ``moving".

To identify those active users with a valid mobility track, we require the following characteristics to be present in each mobility track: (1) a location of home (see below for the inference of home position); (2) records of mobility in all 7 days; (3) a location outside home recorded in at least one time window per day; (4) at most 30 ``moving" time windows among all the 96 time windows per day; (5) less than 24 time windows at home among the 60 time windows per day during the daytime which we define as the period between 6am to 9pm. Finally, 702,477 mobility tracks satisfy the above requirements and were used in our simulations.


\textbf{Inference of users' home, workplace and profession.}
To analyze how the virus transmits inside and outside home, we define the home location of each individual to be the location he/she stayed the longest time between 8pm to 7am the next morning during weekdays, and with at least 6 hours of stay. Since 4G communications from mobile phones sometimes switch between nearby base stations even if the phones are stationary, we substitute the overnight locations of a user which were within 250m from his/her inferred home location by his/her inferred home location, following the definition in~\cite{toole2015path}. Similarly, we define the workplace for each individual to be the location he/she stayed the longest time between 7am to 8pm on the same day during weekdays, and with at least 3 hours of stay.

Based on the workplace, we can first group users into 13 profession categories which include catering staff, village workers, university staff, transportation staff, hotel staff, urban residential workers, industrial and corporate workers, retailers, suburban workers, entertainment staff, medical practitioners, school students and civil servants. We remark that urban residential workers, suburban workers and village workers include individuals whose workplace locations are identified in urban residential areas, suburban areas and villages respectively, but the exact nature of the location is not known. On the other hand,
entertainment staff include providers of entertainment and accessory services such as karaoke, hair salon, public bathrooms, car maintenance, etc. Users whose workplace cannot be identified using the above criteria are first considered as ``freelancers". Some of the freelancers are then considered as retirees as we will describe below. We also remark that although our inference of users' home, workplace, profession and age group may not be perfect, such inference from real data would still lead to much more realistic simulations compared to agent-based simulation studies of which the properties of individual agents are all modeled.

\textbf{Inference of users' age.}
Based on the $7^{th}$ official Census  of Shijiazhuang in 2022, we group users into three age groups, namely (1) under 15, (2) 15 to 60, and (3) above 60. To infer the age group of individual users based on their mobility tracks, we first classify users with profession ``school students" as under 15. On the other hand, to identify elderly users, we calculate the daily total traveling distance and the radius of gyration for all recorded users as shown in Table~\ref{tab_distance}, and for those ``freelancers" with both of these measures ranked in the bottom $15\%$ percentile, they are re-classified as ``retirees" and are included in the age group with age above 60. Finally, the population in the three age groups in each district is consistent to the statistics given by the $7^{th}$ Census of Shijiazhuang.

\textbf{Modeling household members.}
Since our dataset only includes the mobility patterns of individual users without other information, to analyze transmissions at home, we have to select the household members of individual users. From the $7^{th}$ Census of Shijiazhuang in 2022, we obtain the statistics on the number of members in a household as shown in Table~\ref{tab_household}. We then randomly group individual users with the same inferred home location in households according to these statistics, with only one requirement that individuals under 15 cannot form a household by themselves.


\clearpage
\noindent \textbf{Acknowledgments.} \\
C.Z. acknowledges the National Natural Science Foundation of China (No. 61703136), the Natural Science Foundation of Hebei (No. F2020205012), the Youth Top Talent Project of Hebei Education Department (NO. BJ2020035). The work by C.H.Y. is supported by the Research Grants Council of the Hong Kong Special Administrative Region,China (Projects No. EdUHK GRF 18301217, and No. GRF 18301119), the Dean's Research Fund of the Faculty of Liberal Arts and Social Sciences (Projects No. FLASS/DRF 04418, No. FLASS/ROP 04396, and No. FLASS/DRF 04624),and the Internal Research Grant (Project No. RG67 2018-2019R R4015 and No. RG31 2020-2021R R4152), The Education University of Hong Kong, Hong Kong Special Administrative Region, China.

\noindent \\ \textbf{Author contributions.} \\
C. Z., C.H.Y. and A.Z. designed the research, C.Z., J.Z. and X.H. performed the experiments, C. Z., C.H.Y. and A.Z. analyzed the data, C.H.Y. and A.Z. wrote the paper.\\

\noindent \textbf{Competing financial interests.} The authors declare no competing financial interests.\\

\noindent \textbf{Data and materials availability.} The raw data are not publicly available to preserve users' privacy under the Mobile Privacy Policy of China. Derived data supporting the findings of this study are available from the corresponding authors upon request. Due to the data security of participants, 7-days trajectory data cannot be shared freely, but are partly available to researchers who sign a confidentiality agreement and meet the criteria for access to confidential data.




\clearpage
\begin{center}
{\large\bfseries Supplementary Information}\\[8pt]
{\large A high-frequency mobility big-data reveals how COVID-19 spread across professions, locations and age groups}\\
\small Chen Zhao, Jialu Zhang, Xiaoyue Hou, Chi Ho Yeung, An Zeng
\end{center}
\section*{Supplementary Note, basic descriptive statistics of the dataset}
We summarize in this section the basic descriptive statistics of the dataset. We show in Fig. S1A the distribution of travelling
distance in the data. Fig. S1B shows the distribution of the radius of gyration of users in the data. Similar to what has been found in the literature~\cite{understand2008gonzalez}, both distributions follow a truncated power-law, indicating that most individuals move in a small area while a small number of individuals tend to travel a long distance.

In Fig. S1C, we show the distribution of resident population of locations (corresponding to the area each cell-tower covers). The distribution follows a power-law form, suggesting that most locations are visited by few people while a few locations are very crowded. In Fig. S1D, we show the distribution of population in different professions. The three professions with the most population are industrial and corporate workers, freelancers, retirees.

\clearpage
\section*{Supplementary Figures}
\begin{figure}[h!]
  \centering
  \includegraphics[width=16 cm]{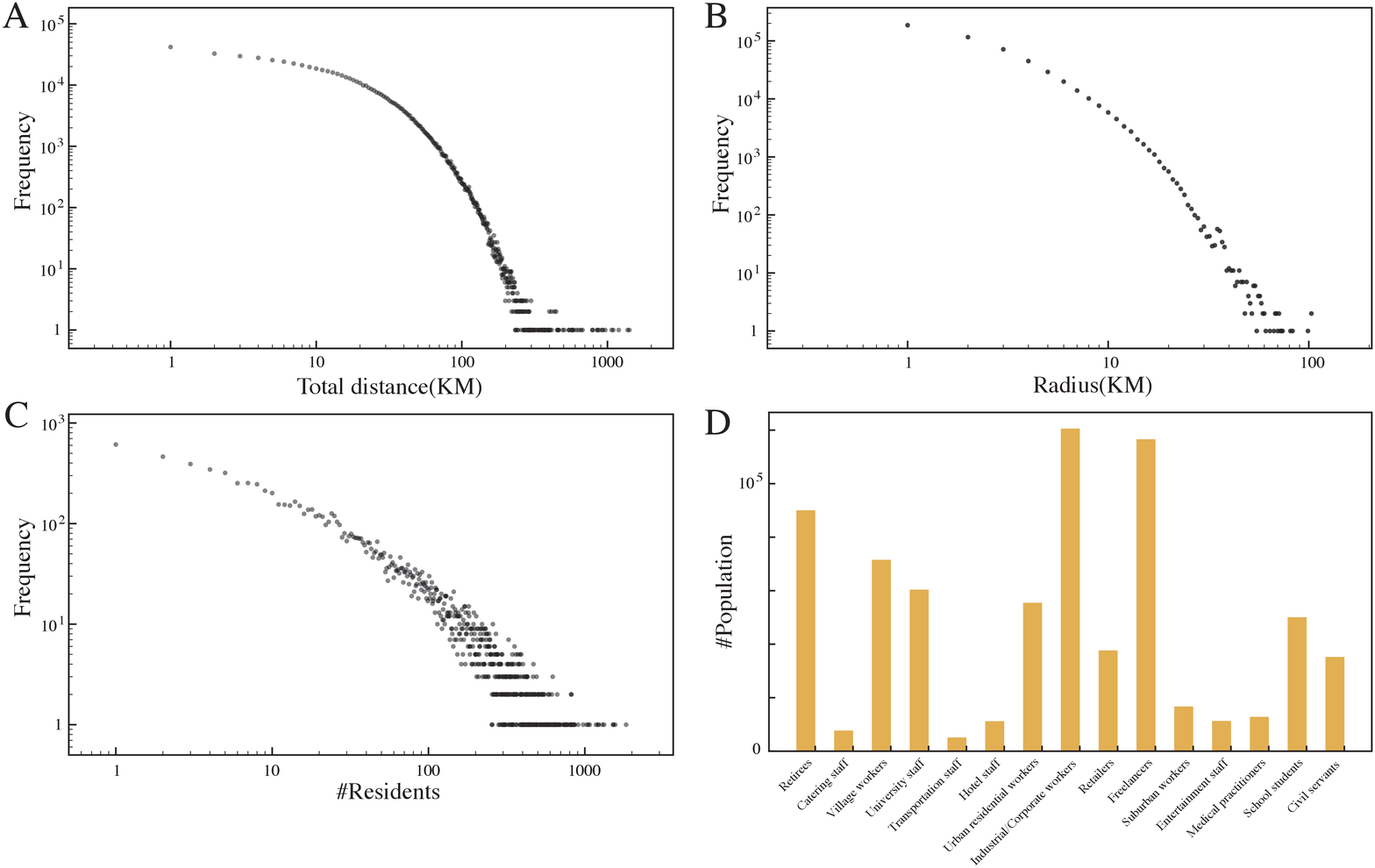}\\
   \textbf{Supplementary Figure S1.} The basic statistics of the data. (a) The distribution of travelling distance in the data. (b) The distribution of the radius of gyration of users in the data. (c) The distribution of resident population of locations (corresponding to the area each cell-tower covers). (d) The distribution of population in different professions. \label{FigS1}
\end{figure}

\clearpage
\begin{figure}[h!]
  \centering
  \includegraphics[width=16 cm]{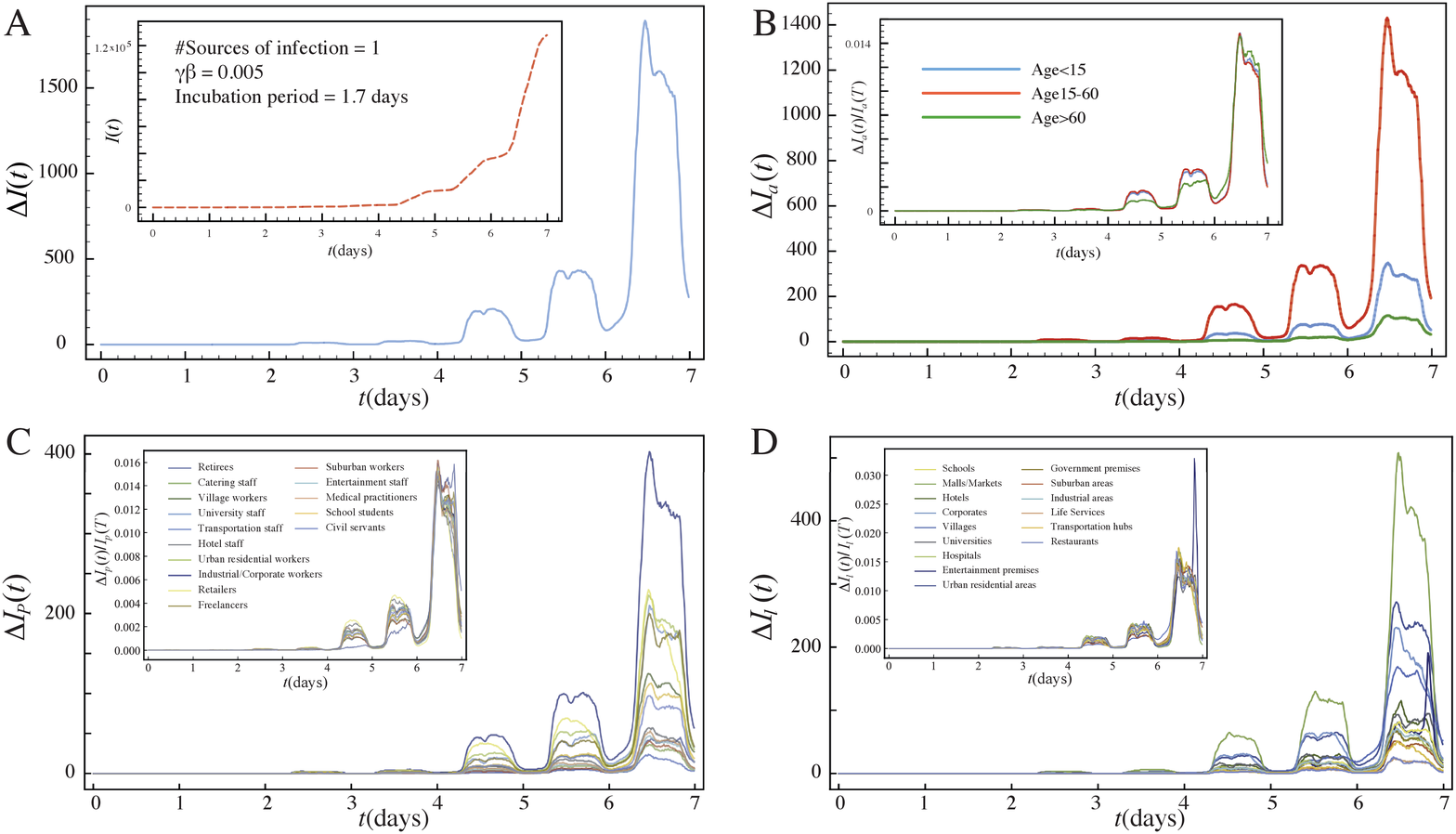}\\
   \textbf{Supplementary Figure S2.} The effect of the number of initial spreaders on the prevalence of the virus. Here, we set only 1 initial spreader, in contrast to the 70 initial spreaders in the main text of the paper. In order to avoid the spreading quickly dies out, we use a relatively larger infection rate as $0.005$ (i.e. 2.5 times larger than the infection rate used in the paper). The rest of the parameters are the same as those used in the paper. (A) Given 1 initial spreader randomly located in the city, the evolution of the size of the infected population per quarter in the city. The inset shows the accumulated infected population in different days. (B) The evolution of the number of infected individuals (per quarter) in different age groups. Inset shows the evolution of the fraction of infected individuals (per quarter) in different age groups. (C) The evolution of the number of infected individuals (per quarter) of different professions in the city. The inset is the evolution of the fraction of infected individuals in different professions. (D) The evolution of the number of infected individuals in different  location categories. The inset is the evolution of the fraction of infected individuals in different location categories. \label{FigS6}
\end{figure}

\clearpage
\begin{figure}[h!]
  \centering
  \includegraphics[width=16 cm]{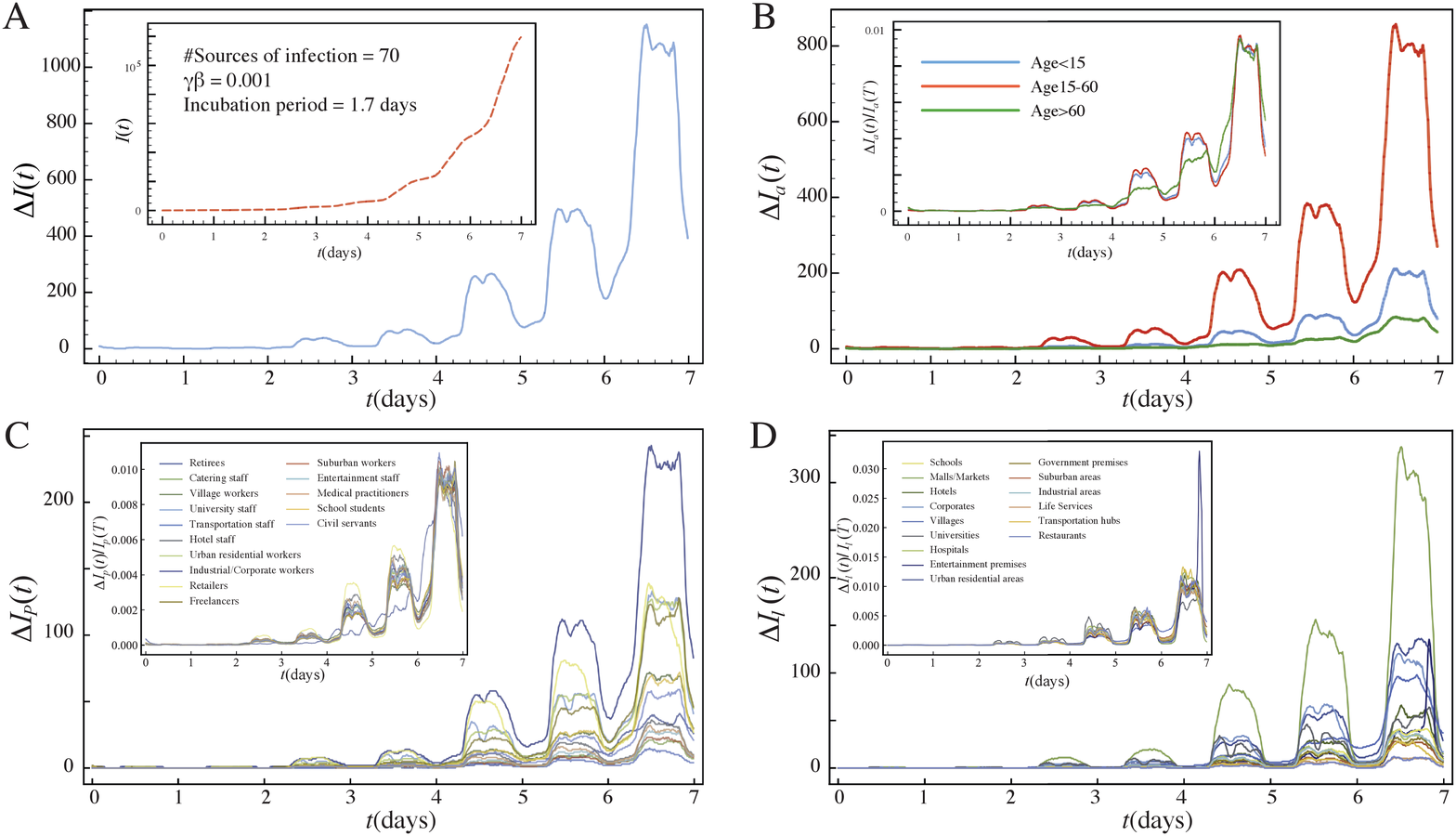}\\
   \textbf{Supplementary Figure S3.} The effect of a smaller infection rate on the  prevalence of the virus. Here, the infection rate outside is $0.001$ and the home infection rate is still 100 times larger, i.e. $0.1$. The rest of the parameters are the same as those used in the paper. (A) Given 70 initial spreaders randomly located in the city, the evolution of the number of infected population per quarter in the city. The inset shows the accumulated infected population in different days. (B) The evolution of the number of infected people (per quarter) of different ages. Inset shows the evolution of the fraction of infected people (per quarter) of different ages. (C) The evolution of the number of infected people (per quarter) of different professions in the city. The inset is the evolution of the fraction of infected people of different professions. (D) The evolution of the number of infected people in different types of locations. The inset is the evolution of the fraction of infected people in different types of locations. \label{FigS4}
\end{figure}

\clearpage
\begin{figure}[h!]
  \centering
  \includegraphics[width=16 cm]{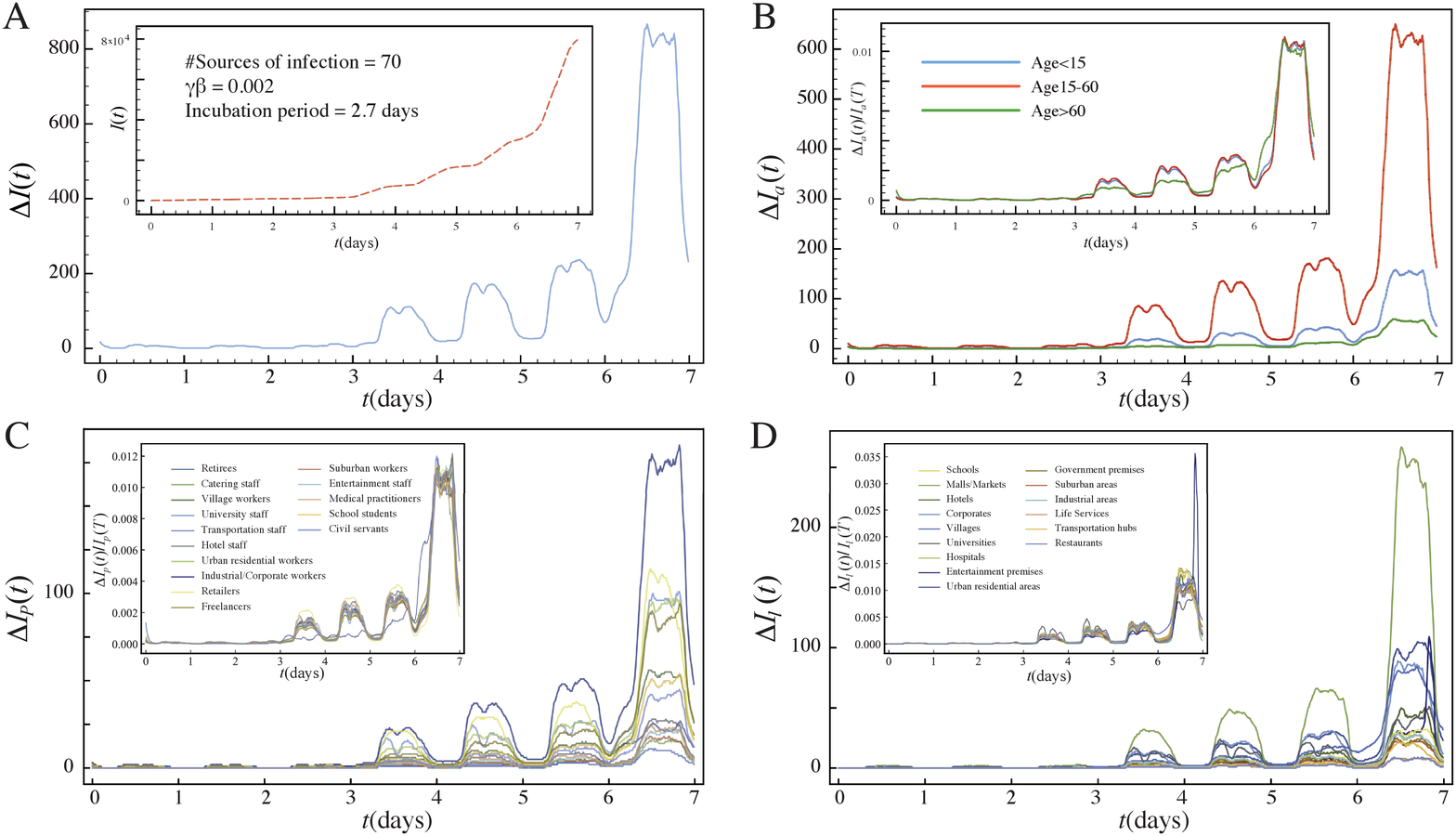}\\
   \textbf{Supplementary Figure S4.} The effect of a longer incubation period on the  prevalence of the virus (an infected individual cannot infect others during the incubation period). Here, the incubation period is set as $2.7$ days, smaller than the value $1.7$ used in the main text of the paper. The rest parameters are the same as those used in the paper. (A) Given 70 initial spreaders randomly located in the city, the evolution of the number of infected population per quarter in the city. The inset shows the accumulated infected population in different days. (B) The evolution of the number of infected people (per quarter) of different ages. Inset shows the evolution of the fraction of infected people (per quarter) of different ages. (C) The evolution of the number of infected people (per quarter) of different professions in the city. The inset is the evolution of the fraction of infected people of different professions. (D) The evolution of the number of infected people in different types of locations. The inset is the evolution of the fraction of infected people in different types of locations. \label{FigS5}
\end{figure}

\clearpage
\begin{figure}[h!]
  \centering
  \includegraphics[width=16 cm]{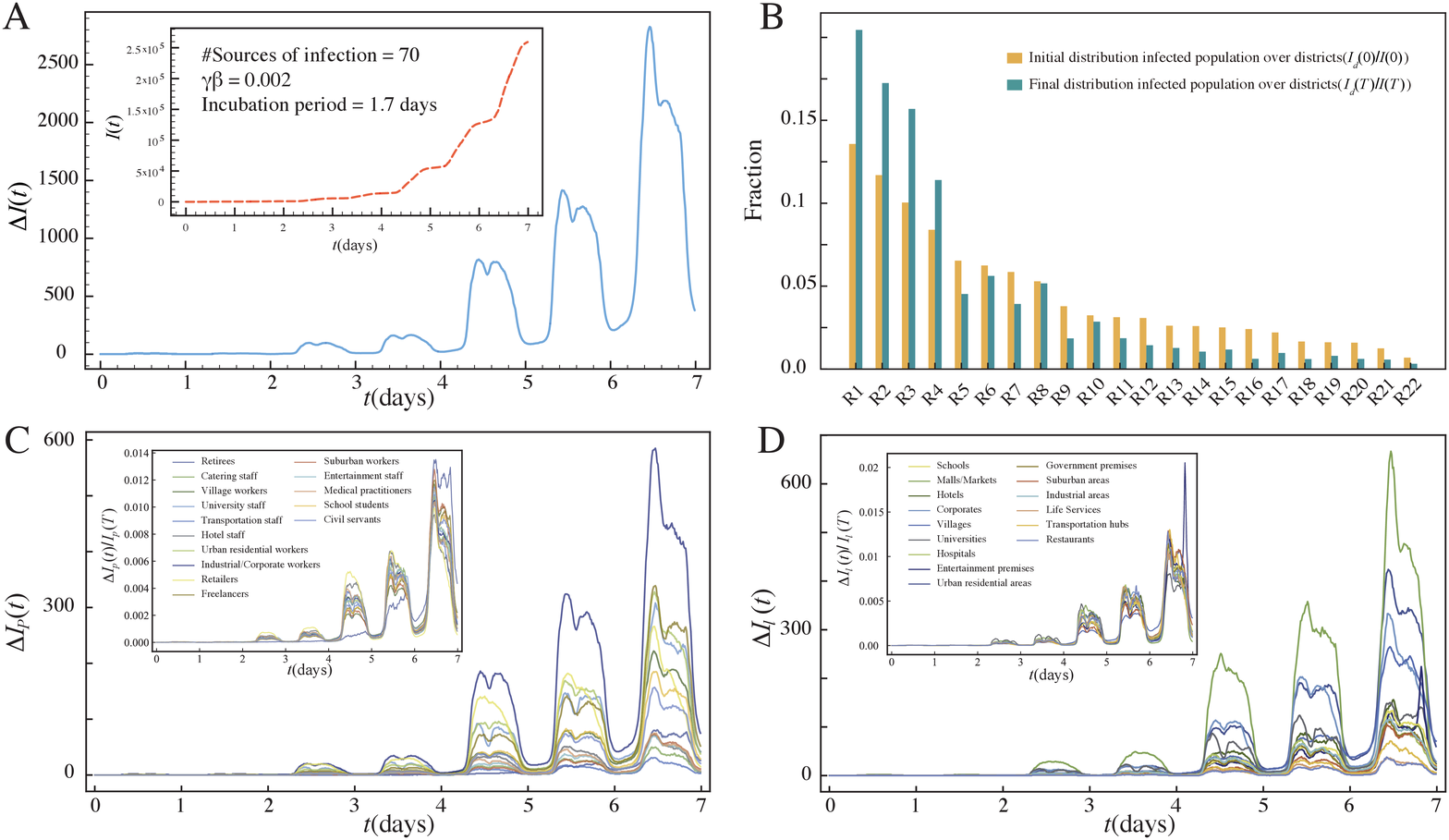}\\
   \textbf{Supplementary Figure S5.} The role of home infection rate on the prevalence of the virus. In this figure, we reduce the home infection rate to $\beta=0.002$ which is the same as the infection rate outside. The rest parameters are the same as those used in the paper. (A) Given 70 initial spreaders randomly located in the city, the evolution of the number of infected population per quarter in the city. A significant periodic infection cycle can be observed, which is caused by the periodic human mobility patterns. The inset shows the accumulated infected population, i.e. $\Delta I(t)$, in different days. (B) The initial and the final distribution of initial individuals in different districts of the city. The districts are shown in the map in Fig. 2 in the main text of the paper. (C) The evolution of the number of infected people (per quarter) of different professions in the city. The inset is the evolution of the fraction of infected people of different professions. (D) The evolution of the number of infected people in different types of locations. The inset is the evolution of the fraction of infected people in different types of locations. \label{FigS2}
\end{figure}

\clearpage
\begin{figure}[h!]
  \centering
  \includegraphics[width=16 cm]{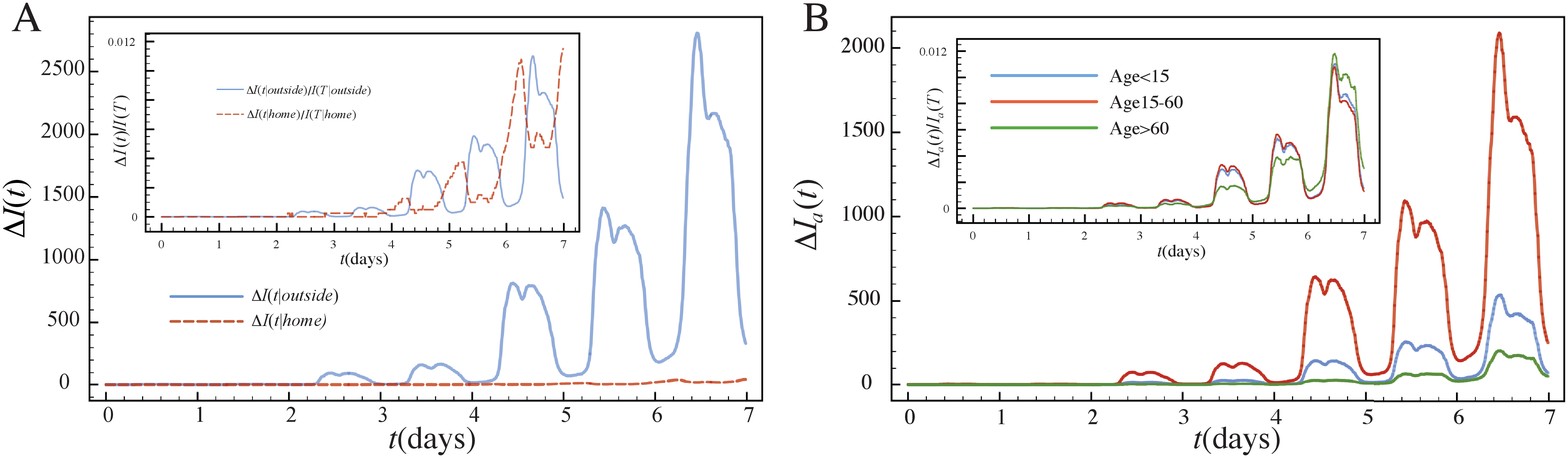}\\
   \textbf{Supplementary Figure S6.} The role of home infection rate on the prevalence of virus. In this figure, we reduce the home infection rate to $\beta=0.002$ which is the same as the infection rate outside. The rest of the parameters are the same as those used in the paper. (A) The evolution of the number of individuals (per quarter) infected at home and the number of individuals (per quarter) infected outside. The inset shows the evolution of the fraction of infected individuals who are infected at home and outside, respectively. (B) The evolution of the number of infected individuals (per quarter) in different age groups. Inset shows the evolution of the fraction of infected individuals (per quarter) in different age groups. \label{FigS3}
\end{figure}

\clearpage
\begin{figure}[h!]
  \centering
  \includegraphics[width=16 cm]{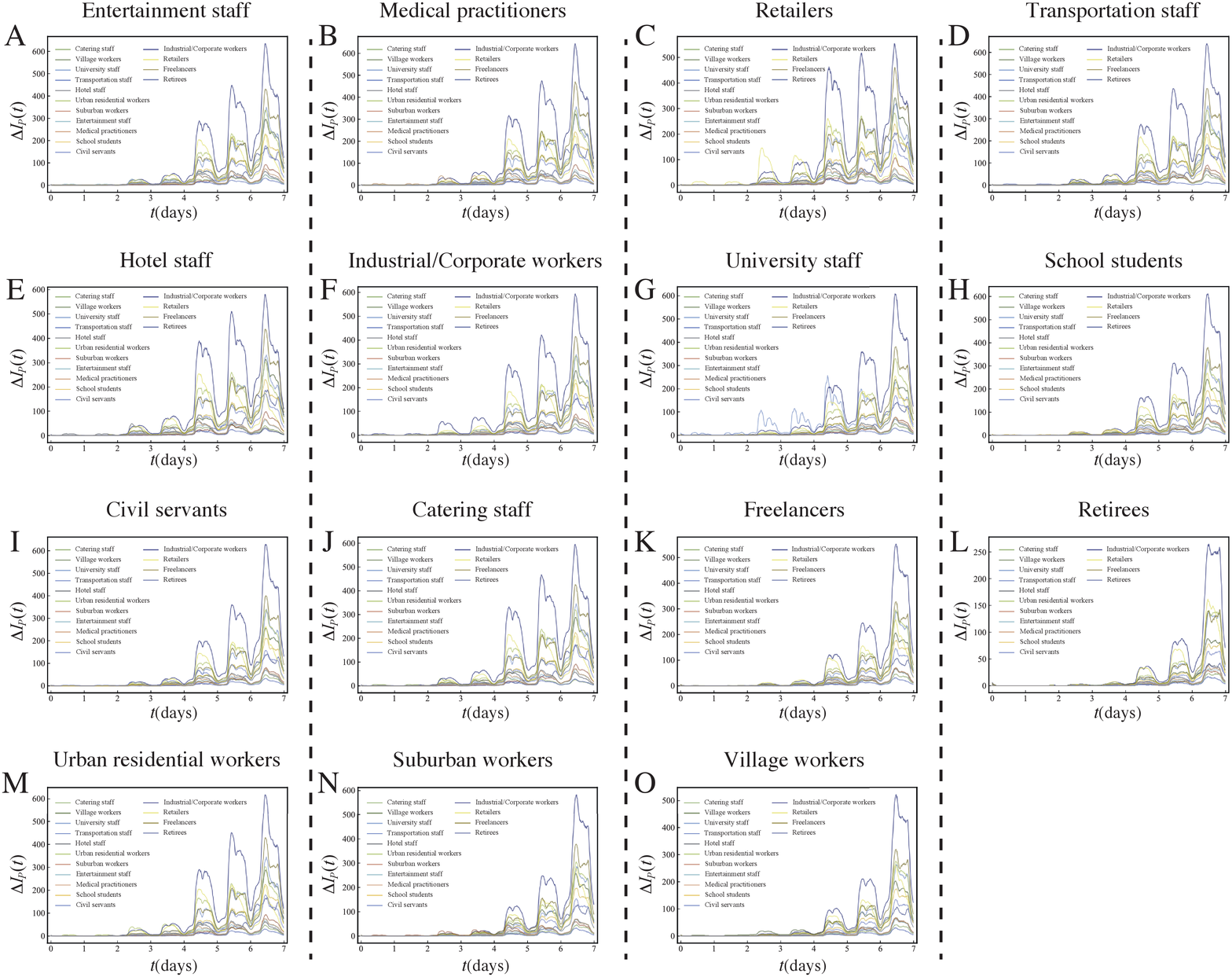}\\
   \textbf{Supplementary Figure S7.} We simulate the spreading results given that the infection starts at individuals of different professions, respectively. 70 initial spreaders are randomly selected from the population of a given profession. The spreading is continued with the intimate contact mechanism for 7 days. All parameters are the same as those used in the paper. The professions are (A) Entertainment staff, (B) Medical practitioner, (C) Retailer, (D) Transportation staff, (E) Hotel staff, (F) Industrial and Corporate Workers, (G) University staff, (H) School student, (I) Civil servant, (J) Catering staff, (K) Freelancer, (L) Retired, (M) Resident, (N) Rural resident, (O) Village resident. In all cases, the significant periodic infection cycle can be observed. However, the spreading initialized from different professions exhibits significant heterogeneity.\label{FigS8}
\end{figure}

\clearpage
\begin{figure}[h!]
  \centering
  \includegraphics[width=16 cm]{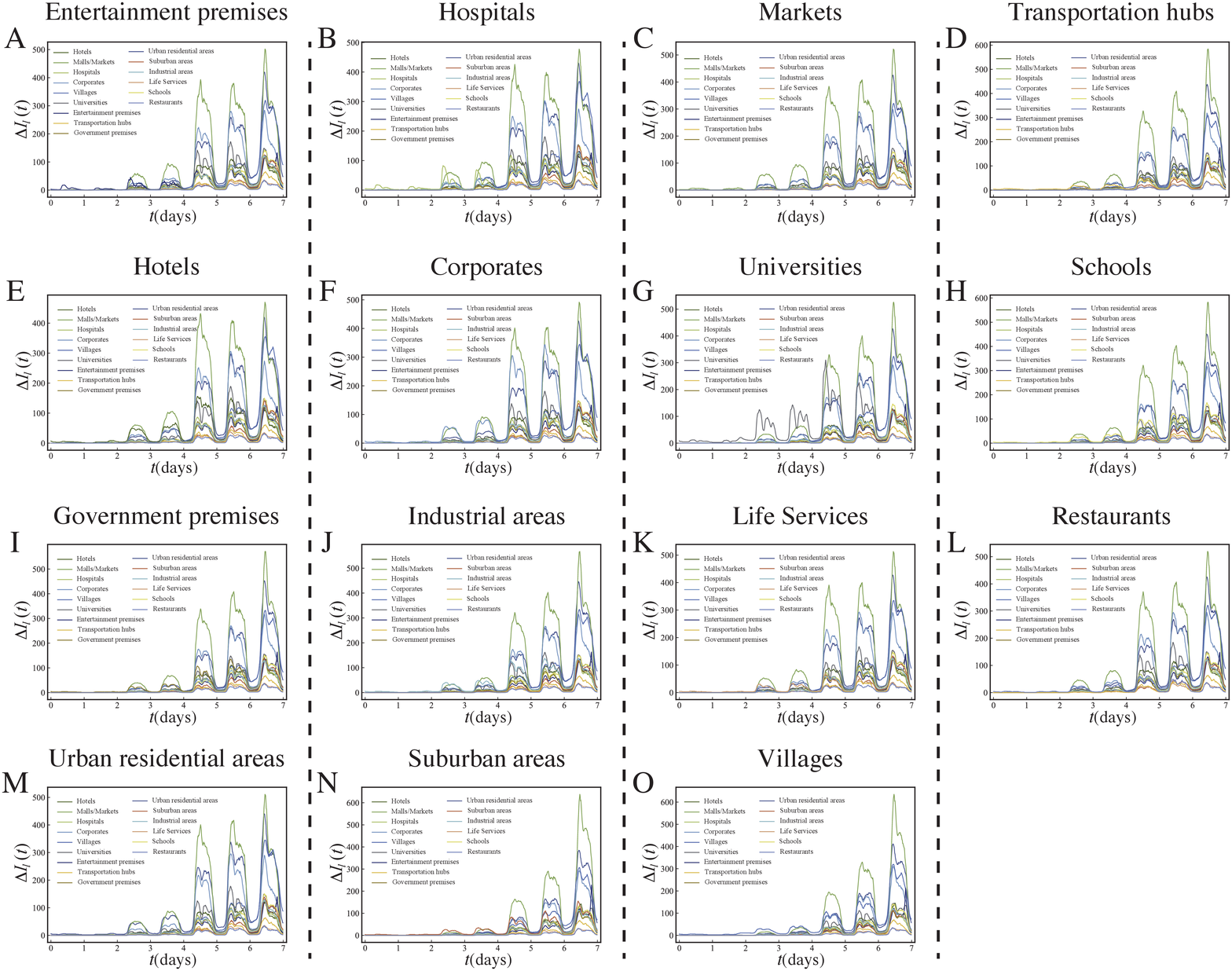}\\
   \textbf{Supplementary Figure S8.} We simulate the spreading results given that the infection starts at locations in different location categories, respectively. The initial spreaders are generated by the environmental infection. Assuming each individual would get infected with an infection rate $0.002$ when visiting the designated location, the environmental infection is terminated when the number of initial spreaders reach 70. The spreading is continued with the intimate contact mechanism for 7 days. All parameters are the same as those used in the paper. The locations designated for environmental infection are (A) Entertainment premises, (B) Hospital, (C) Market, (D) Transportation hub, (E) Hotel, (F) Corporate, (G) University, (H) School, (I) Government premises, (J) Industrial area, (K) Life Services, (L) Restaurant, (M) Urban Residential area, (N) Suburban area. (O) Village. In all cases, the significant periodic infection cycle can be observed. However, the spreading initialized from different locations exhibits significant heterogeneity. In particular, the spreading evolution initialized from suburban areas or villages is different from that started in other locations, possibly due to the different human mobility patterns in urban and suburban area~\cite{switchover2021odor}.\label{FigS7}
\end{figure}

\clearpage
\begin{figure}[h!]
  \centering
  \includegraphics[width=16 cm]{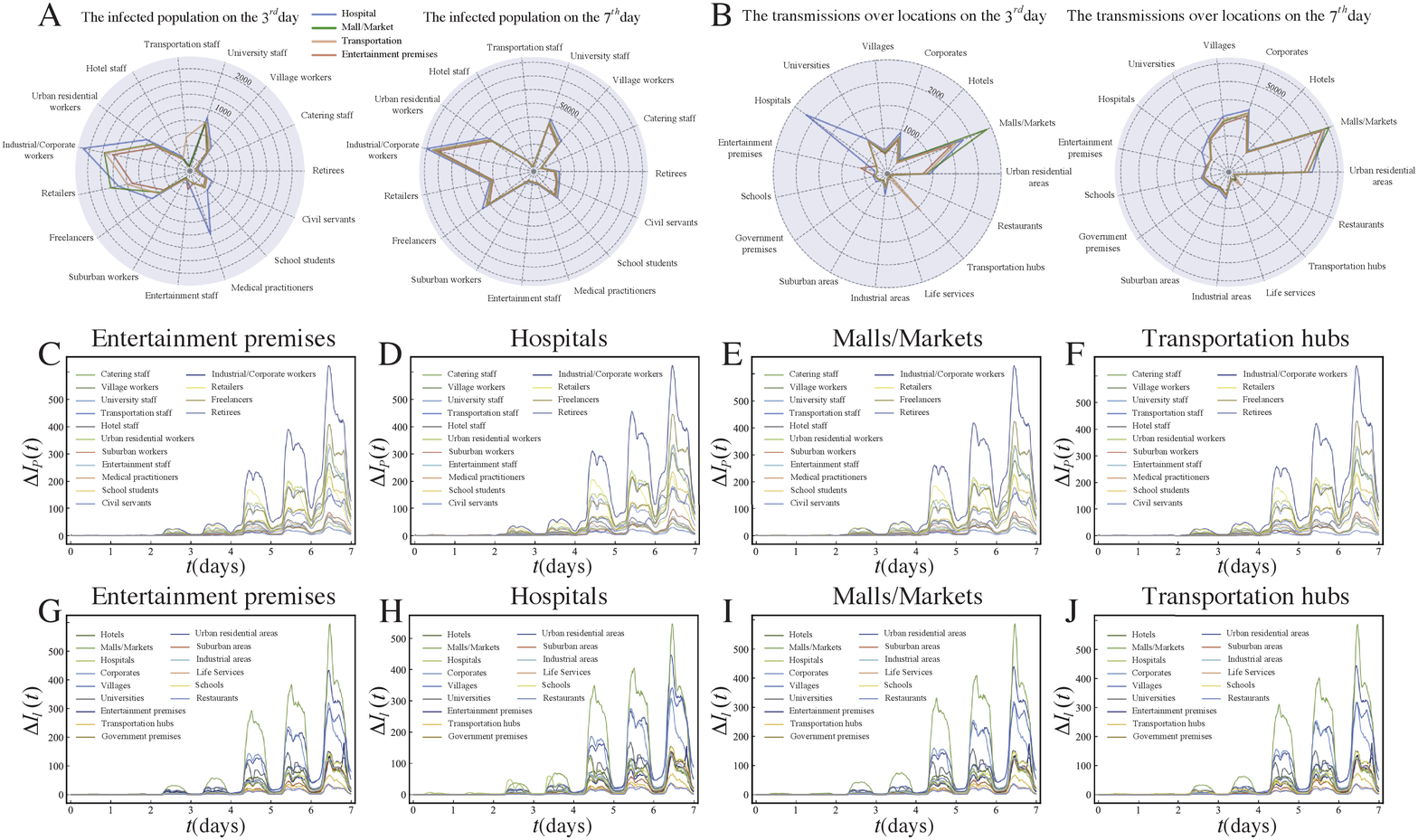}\\
   \textbf{Supplementary Figure S9.} The simulation results where transmissions through both contact and environment are considered. In this case, we first randomly select 70 individuals as the initial infected group. We then consider four different types of locations for environmental spreading, respectively. If a location is a source of environmental spreading, any individual who visits it would have a probability $0.002$ to be infected. We consider respectively four different types of locations as the source, namely entertainment, hospital, market and transportation hub. For the location category ``entertainment premises¡¯¡¯, we select 12 major locations labeled by this location category. For transportation type of locations, we select 20 major locations marked by this location category (mostly train stations, bus stations and airports). For hospital type of locations, we select 31 major locations labeled by this location category (mostly hospitals and institutions of disease control. For  location category ``markets¡¯¡¯, we select 17 major locations marked by this location catergory (mostly supermarkets). (A) The radar map showing the distribution of infection across professions until the third day and the last day. (B) The radar map showing the distribution of infected locations until the third day and the last day. (C-F) The evolution of the number of infected population of different professions per quarter in the city. (G-J) The evolution of the number of infected population in different locations per quarter in the city. \label{FigS9}
\end{figure}

\clearpage
\section*{Supplementary Tables}

\begin{table}[h!]
\caption{The population in different districts of Shijiazhuang city, together with the percentage of population in different age groups in each district.}
\begin{tabular}{cccccc}
  \hline
  District & Name &Population &Age$\in[0,14]$ &Age$\in[15,60]$ &Age beyond $60$\\
  \hline
  R & Shijiazhuang (whole city) &10,640,458 & 19.5\% & 62.5\% & 18.0\% \\
  R1 &Yuhua district &771,255 & 17.1\% & 69.1\% & 13.8\% \\
  R2 &Qiaoxi district &979,646 & 15.9\% & 68.6\% & 15.5\% \\
  R3 &Changan district &1,059,572 & 17.6\% &67.2\% &15.2\% \\
  R4 &Xinhua district &802,057 & 17.0\% & 68.0\% & 15.0\% \\
  R5 &Gaocheng district &741,068 & 21.0\% & 59.5\% & 19.5\% \\
  R6 &Zhengding district &549,321 & 19.9\% & 62.3\% & 17.8\%\\
  R7 &Xinji district &594,628 &16.1\% & 57.8\% & 26.1\% \\
  R8 &Luquan district &588,279 & 19.4\% &63.6\% & 17.0\% \\
  R9 &Jinzhou district &507,959 & 20.5\% & 57.2\% &22.3\% \\
  R10 &Ruancheng district &378,689 & 19.9\% & 63.2\% & 16.9\% \\
  R11 &Xinle district &478,529 & 22.5\% & 60.4\% & 17.1\% \\
  R12 &Wuji district &451,377 & 21.4\% & 57.6\% & 21.0\% \\
  R13 &Pinshan district &423,333 & 21.1\% & 56.5\% & 22.4\% \\
  R14 &Zhao district &505,366 & 21.5\% & 57.9\% & 20.6\% \\
  R15 &Yuanshi district &392,710 & 21.0\% & 60.0\% & 19.0\% \\
  R16 &Xingtang district &376,627 & 22.3\% & 56.1\% & 21.6\% \\
  R17 &Lingshou district &309,121 & 20.3\% & 59.2\% & 20.5\% \\
  R18 &Zanhuang district &242,549 & 27.0\% & 54.7\% & 18.3\% \\
  R19 &Jingxing district &250,989 &16.7\% & 57.7\% & 25.6\% \\
  R20 &Shenzhe district &215,806 & 18.4\% & 55.9\% & 25.7\% \\
  R21 &Gaoyi district &178,368 & 23.4\% & 55.8\% & 20.8\% \\
  R22 &Jingxingkuang district &77,015 & 15.8\% & 61.0\% & 23.2\% \\
  \hline
\end{tabular}
\end{table}

\clearpage
\begin{table}[h!]
\caption{The distribution of family size. The data is the basis for grouping individuals into families which are further used for the home infection in the spreading model. }
\begin{tabular}{ccccccccccccc}
  \hline
  Family member && 1 &&2 &&3  &&4  &&5  &&6 \\
  \hline
  ratio && 12\% &&27\% && 21\% && 20\% && 10\% && 9\%\\
  \hline
\end{tabular}
\end{table}

\end{document}